# Finding COVID-19 from Chest X-rays using Deep Learning on a Small Dataset

Lawrence O. Hall, Rahul Paul, Dmitry B. Goldgof, and Gregory M. Goldgof[1]
Department of Computer Science and Engineering, ENG060
University of South Florida
Tampa, FL 33620
*{lohall,goldgof,rahulp@mail.usf.edu}*
[1] Dept. of Laboratory Medicine
University of California, San Francisco
gregory.goldgof@ucsf.edu

**Abstract**

Testing for COVID-19 has been unable to keep up with the demand. Further, the false negative rate is projected to be as high as 30% and test results can take some time to obtain. X-ray machines are widely available and provide images for diagnosis quickly. This paper explores how useful chest X-ray images can be in diagnosing COVID-19 disease. We have obtained 135 chest X-rays of COVID-19 and 320 chest X-rays of viral and bacterial pneumonia. A pre-trained deep convolutional neural network, Resnet50 was tuned on 102 COVID-19 cases and 102 other pneumonia cases in a 10-fold cross validation. The results were an overall accuracy of 89.2% with a COVID-19 true positive rate of 0.8039 and an AUC of 0.95. Pre-trained Resnet50 and VGG16 plus our own small CNN were tuned or trained on a balanced set of COVID-19 and pneumonia chest X-rays. An ensemble of the three types of CNN classifiers was applied to a test set of 33 unseen COVID-19 and 218 pneumonia cases. The overall accuracy was 91.24% with the true positive rate for COVID-19 of 0.7879 with 6.88% false positives for a true negative rate of 0.9312 and AUC of 0.94. This preliminary study has flaws, most critically a lack of information about where in the disease process the COVID-19 cases were and the small data set size. More COVID-19 case images at good resolution will enable a better answer to the question of how useful chest X-rays can be for diagnosing COVID-19. [1]

# 1 Introduction

Identifying who has the COVID-19 virus is critical for controlling its spread. With testing ability limited in the US and other countries, some tests requiring significant time to produce results (days), and a projected up to 30% false negative rate, other timely approaches to diagnosis are worthy of investigation [1]. Gold standard testing currently is a RNA-based assay using nasopharyngeal swabs. The test is uncomfortable and invasive, and there are already nationwide shortages of swabs, viral transport media, and reagents needed for testing. Chest X-rays (CXR) can be used to give relatively immediate diagnostic information. X-ray machines are widely available and scans are relatively low cost and are ubiquitous in both emergency and hospital settings, where interpretation is often performed without expert radiologists. In addition, X-rays can be performed without increased risk of aerosolizing the pathogen, unlike laboratory tests that involve probing the patient's respiratory system. The X-rays may also enable the triage of patients into highest risk, high risk and lower risk of further complications, as well as to indicate the severity of disease at one or more time points. The focus of this preliminary paper is to determine whether chest X-rays may help with diagnosis of COVID-19.

There has been some other recent work on using chest X-rays for COVID-19 diagnosis that also indicate there may be some level of utility [2, 3, 4]. The results, while also on relatively small datasets, have good sensitivity to COVID-19.

To see if chest X-rays may be viable as method for diagnosis, we worked with 135 chest X-rays of patients diagnosed with COVID-19 and a set of 320 chest X-rays of patients diagnosed with either viral pneumonia or bacterial pneumonia that predates the emergence of COVID-19. It is thought the other types of pneumonia would be most difficult to differentiate from COVID-19. Transfer learning using a pre-trained deep convolutional neural network (CNN) has been done to see if it may be possible to diagnose COVID-19 from CXR. An issue with our data is that the stage of disease is unknown for the COVID-19 group, only that they had abnormal chest X-rays. The COVID-19 images are also mostly reduced quality jpeg. However, from our small data set it can be seen that it *may* be feasible to use CXR for diagnosis, especially if there are breathing issues.

---

[1]Note an earlier version of this work inadvertently used chest X-rays of viral and bacterial pneumonia that came from a dataset of children under 5 years old and those results should be ignored. This version also fixes a problem with combing our Snapshot learning results.

## 2 Data

There were 135 COVID-19 cases obtained from 3 sources
(A: https://github.com/ieee8023/covid-chestxray-dataset (92 cases) [5],
B: https://radiopaedia.org/search?utf8=%E2%9C%93&q=covid&scope=all&lang=us (33 cases) and C: http://www.sirm.org/en/ (10 cases)). The 92 in A and the 10 available in B on March 23, 2020 were used in training. The images were in jpeg format and had a typical quality 0.9 meaning some lossy compression had unfortunately been applied. The cases in C plus the 23 from after March 23, 2020 in B made up testing. The 320 pneumonia cases come from https://www.kaggle.com/nih-chest-xrays/data. These cases were in png format and did not have reduced resolution to our knowledge. To create a balanced training set, we used the first 102 COVID-19 cases acquired and randomly chose 102 other pneumonia cases to get 2/3 being viral pneumonia cases and 1/3 bacterial pneumonia cases.

## 3 Experimental setup

Both the VGG16 CNN [6, 7] and Resnet50 [8] which were trained on color camera images from Imagenet were used in doing transfer learning. VGG16 has 16 layers and we used the convolutional layers without change. For Resnet50, we also used the convolutional layers without change. Both have 3 channels for RGB, and X-rays are grayscale, the same image scaled to the size of 224x224 was provided to each channel. For VGG there are 13 convolutional layers, each uses a small filter of size 3x3 with 1 stride and padding of 1 to extract features. Five max pooling layers with a 2x2 window and stride of 2 are used after each block of the convolutional layers. In our model, the last layers of VGG16 were removed and replaced by the trainable part which consisted of global average pooling, followed by a fully connected layer of 64 units with dropout and finally a classification layer with sigmoid output. The same procedure was followed for Resnet50. The cyclic learning rate (https://github.com/bckenstler/CLR) was used with Base learning rate= 0.0001 and Max learning rate= 0.001. RMSprop was used for learning with binary-cross-entropy as the loss function.

With this small data set, a 10-fold cross validation was done using Resnet50 because it had the highest accuracy on our unseen data set. The only augmentation done was horizontal flipping. As we did not have enough data to set aside a validation set for parameter tuning or stopping, we chose to use a Snapshot ensemble [9] of 5 classifiers. When the training accuracy curve reached > 90% accuracy, we chose 7 models from the stored weights of models starting with the last epoch with a gap of 10 to create the snapshot ensemble. The outputs of the 7 models were averaged to determine the class.

We also tried a small experiment on unseen data consisting of the later collected COVID-19 data (33 cases) and the rest of the pneumonia cases (208). For this, we did the same type of transfer learning with Resnet50, and VGG16 with a different data setup. Of the 102 images in each class, 18 of each class were used in a validation set (36 images) and then horizontal flipping was done for augmentation of the rest which made up the training data. A Snapshot ensemble of 7 models was created as follows. When the validation accuracy was > 0.8, starting from that epoch, 7 models were extracted for voting with a gap of 10 epochs between them. We used a small CNN trained just on this data that we have used before [10] the same training approach was used to get 7 snapshot models. The 21 models were used to predict the unseen test set of 241 cases by averaging the outputs and selected the class with the highest average.

## 4   Results

A 10-fold cross validation was done to get an estimate of the feasibility of using chest X-rays to diagnose COVID-19. The initial results, if they can be improved on bigger more diverse data sets, show some promise. The overall accuracy was 89.2% with 80.39% of the COVID cases correctly identified and other pneumonia correctly identified 101/102 for specificity/true negative rate (TNR) TNR= 0.99. Consequently, there was 1 false positive of other pneumonia. Our Overall AUC was 0.95.

We had a number of unseen pneumonia cases (208 cases) and acquired new COVID-19 cases (33) not in training. Since, the COVID-19 cases had reduced resolution we converted all pneumonia cases into jpeg format and reduced the to a quality of 0.9. This was done to try to ensure that the sharpness of an image was not unduly influencing predictions. With a Snapshot ensemble of 21 models from Resnet50, VGG16 and our own small CNN created using all 204 images for training 26 of 33 COVID-19 cases were correctly identified. The true negative rate was 93.12% meaning about 6.0% false positives. The AUC was 0.94.

# 5    Discussion and Conclusions

There has been work on diagnosing COVID-19 from computed tomography (CT) scans [3] which showed promise and is discussed below. This work focuses on CXRs which are simpler to obtain, more widely available worldwide and cheaper to obtain, though provide less information that CT. Our current work has about a 6% false positive rate that may be reduced by biasing the training data to include more non COVID-19 cases and 83.3 to 96% true positive rate. Though one does not want to reduce the true positive rate, the ratio of examples could be modified to reduce the FPR, but would require experiments that are best facilitated by more CXR data. Full resolution of CXR COVID-19 data would be most useful. Given the existing chest CT's of COVID-19 that are available, the option to create pseudo chest X-rays from them is an, albeit unusual, option for adding to our data set of COVID-19 cases.

A recent paper looked at 4,356 chest CT exams from 3,322 patients [3] with 1,296 (30%) exams for COVID-19, 1,735 (40%) for pneumonia and 1,325 (30%) for non-pneumonia. So, a good sized data set from which they concluded that CT had diagnostic utility for COVID-19 diagnosis. An AUC=0.96 was obtained for COVID-ID identification. They used deep learning to extract a 3D view of the lungs and then a Resnet50 [8] based model to do the diagnosis. The results suggest CT can be useful for COVID-19 diagnosis, so maybe CXRs, despite the reduced information they contain, may also be useful.

Several very recent papers have also shown results with just chest X-rays for COVID-19. In [4] 1,427 CXR images were used which consisted of 224 images with confirmed COVID-19, 700 images with confirmed common pneumonia, and 504 images that were normal. They did transfer learning and looked at several models finding VGG19 to be the best. In a 10-fold cross validation they found true positives - TP=208, false negatives - FN=16, true negatives - TN=1189, false positives - FP=15. While it appears that some tuning may have been done in the cross validation making for slightly optimistic results, the results are positive. In [2] they developed a deep learning anomaly detection approach for screening from chest X-rays. Anomaly detection is a more challenging problem formulation, as it is one class classification. They did correctly classify 96% of 100 chest X-rays from 70 COVID-19 patients. On 1,431 CXR from 1,008 pneumonia patients they had 76.5% accuracy. So, they had quite a few more false positives than we did, yet still there is differentiation shown. In [11] transfer learning was done to identify CXRs as pneumonia or normal. They obtained 96% accuracy with a Resnet50 model on a data set of 4,273 pneumonia and 1,583 normal with 60% of the data used for training and 40% used for testing. This again suggests there is a possibility of effectively using deep learning on chest X-rays for disease detection. In [12] a Hong Kong study, it was shown chest X-ray abnormalities in COVID-19 match those seen in CT, and show bilateral peripheral consolidation. They looked at 64 CXR cases and did find a 69% sensitivity on a baseline scan, which was lower than the non-imaging test they applied (91%),

Our data set only includes patients with CXR findings, and as a result our algorithm is unable to detect disease in patients who do not have human-observable CXR findings on presentation (31% as suggested by the Hong Kong study). A machine learning approach may be able to identify abnormalities in patients who have normal appearing CXR findings to a human interpreter, however, this hypothesis can only be explored after development of a dataset of CXRs from COVID-19 positive patients that also includes images that have been read as negative by an MD. Importantly, the Hong Kong study also suggests the possibility of using CXRs to identify COVID-19 positive patients with false negative laboratory tests, a significant problem facing physicians with current estimates for nasopharyngeal swabs as high as 30%, and even higher for oropharyngeal swabs.

In work on lung cancer screening from low dose CT images [10], small data sets have been used to build effective ensembles of CNN's that can predict lung nodules that become malignant between 1 and 2 years after imaging with an AUC=0.94. Hence, a next step is to build CNN type models without transfer learning. It is also possible to create transfer learning ensembles and add those into the mix of classifiers whose results will be combined.

This preliminary work with just 102 COVID-19 chest X-rays used in training indicates it may, with more data, be possible to build deep neural network models for accurate diagnosis. In a 10-fold cross validation experiment with 102 cases of COVID-19 and 102 cases of other pneumonia, overall accuracy was 89.2% with 80.39% of COVID-19 cases correctly identified. Using a model built from all training data applied to 33 unseen COVID-19 cases and 208 viral and bacterial pneumonia cases the true positive rate was 78.79% with a true negative rate of 93.12%. Among the many limitations of this work is the small number of cases available, the lack of information on what stage of disease the chest X-rays are from, the lack of information on outcomes, the reduced resolution of the COVID-19 images and the lack of information on the types of chest X-rays most likely to be confused. Nevertheless, the current results indicate that a deep learned model applied to chest X-rays may, with more data, be of some help in diagnosing COVID-19. The collection of more COVID-19 images is the most important task towards effectively answering the question of whether chest X-rays can be a truly useful tool for fast, accurate, relatively inexpensive diagnosis of the disease. [2]

---

[2]Data and code are available in a Github repository at: https://github.com/hellorp1990/Covid-19-USF